\documentclass[dvips]{acta}
\usepackage{supertabular,lscape,epsfig}
\usepackage{amssymb}
\usepackage{amsmath}
\usepackage{graphicx}
\usepackage{xcolor}

\SetPages{0}{0}

\SetVol{58}{2008}

\begin{document}

\begin{Titlepage}
\Title{Light curve modeling of four short-period W UMa binaries}

\Author{{\v C}~e~k~i, A.~and~~L~a~t~k~o~v~i~{\' c}, O.}
{Astronomical Observatory, Volgina 7, 11060 Belgrade, Serbia\\
e-mail: atila@aob.rs}

\Received{Month Day, Year}
\end{Titlepage}

% Abstract

\Abstract{We use extensive model grids to estimate the global parameters of four partially-eclipsing W UMa contact binaries near the period cutoff. All four systems consist of K-type main sequence primaries and M-type secondaries that appear undersized and underluminous for their masses because of the energy transfer through the common envelope. Three of the four stars exhibit light curve asymmetry that is explained in terms of magnetic activity and modeled with dark spots. We discuss the reliability of the photometric mass ratios and derived absolute parameters in context of total or partial eclipses and compare them with a sample of totally-eclipsing short-period W UMa systems from the literature.}
{binaries (including multiple): close -- binaries: eclipsing -- stars: fundamental parameters -- stars: individual: 
2MASS J02272637+1156494, 
1SWASP J040615.79-425002.3, 
1SWASP J121906.35-240056.9, 
2MASS J2326101-2941470}

% Sections

\section{Introduction}

In a previous paper (Latkovi{\' c} \& {\v C}eki 2021; hereafter, P1), we investigated six of the 29 southern eclipsing binaries with short periods and W UMa-like light curves observed by Koen et al. (2016). We estimated their photometric mass ratios using the ``q-search'' method. In general, the mass ratio of a binary can only be established from radial velocity measurements (the spectroscopic mass ratio); but in contact binaries, the components share a common envelope defined by a single Roche surface, whose size is uniquely determined by the mass ratio. As sizes of the components influence the shape of the light curve, it is in principle possible to estimate the mass ratios of contact binaries even when radial velocities are not available. The six targets of P1 had either a readily recognizable total eclipse, or could only be matched with totally eclipsing models. The totality aids the analysis of a contact binary by decoupling the mass ratio from the orbital inclination, which has a similar effect on the shape of the light curve, so that the two parameters are generally correlated. The total eclipse breaks the correlation by constraining the inclination to a small range close to 90$^{\circ}$. For a detailed discussion of how totality (or the lack thereof) affects the reliability of photometric mass ratios in contact binaries, see e.g. Latkovi{\' c}, {\v C}eki \& Lazarevi{\' c} (2021). When the eclipses are partial, the photometric mass ratio is far less reliable than the spectroscopic one.

In the present study, we go on to examine four more W UMa binaries from the same set of observations made by by Koen et al. (2016): 2MASS J02272637+1156494 (hereafter J022726), 1SWASP J040615.79-425002.3 (hereafter J040615), 1SWASP J121906.35-240056.9 (hereafter J121906), and 2MASS J2326101-2941470 (hereafter J232610). Among these stars, only J022726 has been studied previously (Liu et al. 2015). For the others, this work is the first published analysis. While they do not display a visible totality in the available observations, we show that their mass ratios and inclinations are constrained to a well-defined region of the parameter space using a high-resolution model grid. This allows us to derive reasonable estimates of their orbital and stellar parameters.

The procedure of finding the photometric mass ratio of a contact binary by modeling of the light curve, where, for a series of fixed mass ratio trial values, one adjusts all the other model parameters and eventually chooses the mass ratio of the best-matching model, is known as the q-search. We performed the q-search for all 29 stars observed by Koen et al. (2016) using a heuristic approach where the other parameters of candidate models at each fixed mass ratio were randomized (the procedure is detailed in P1). This gave us an overview of the parameter space so that we were reasonably confident that our solutions were not local minimums. The current study was motivated by the appearance of the resulting q-search curves (where one plots the reduced $\chi^2$, i.e. $\chi_{\nu}^2$, or some other goodness-of-fit indicator, against the mass ratio); namely, the four stars that we investigate now have similar q-search curves as the six studied in P1, despite the apparent lack of totality.

For these new targets, we redo the q-search in two dimensions, with the inclination on the other axis, and show that the region containing all the best-fitting models is well-defined. The procedure is detailed in Section 3. Starting from the best-fitting models found during the q-search, we perform detailed modeling described in Section 4, and compare the results with a selection of similar objects in Section 5.

\section{Data Preparation}

We prepared the light curves of our four stars the same way as in P1. Their general characteristics are listed in Table 1. The coordinates and magnitudes are taken from the Simbad database, and the orbital periods from Norton et al. (2011). The available data allowed us to measure a single pair of eclipse times for each star. This was done by fitting a low-order polynomial through the minimum. We used the eclipse times for the deeper minimum (listed in Table 2 as $\rm T_I$) and the periods in Table 1 to calculate the orbital phases.

\MakeTable{llccccc}{\textwidth}{The main characteristics of the studied stars.}
{
  \hline
  Nickname & ID & RA & DEC & $m_V$ & Period [d] \\
  \hline
  J022726 & 2MASS  J02272637+1156494   & 02 27 26.38 & +11 56 49.45 & 15.11 & 0.21095 \\
  J040615 & 1SWASP J040615.79-425002.3 & 04 06 15.83 & -42 50 02.33 & 14.14 & 0.22234 \\
  J121906 & 1SWASP J121906.35-240056.9 & 12 19 06.33 & -24 00 56.96 & 15.25 & 0.22637 \\
  J232610 & 2MASS  J23261012-2941470   & 23 26 10.12 & -29 41 47.08 & 13.58 & 0.23012 \\
  \hline
  \multicolumn{7}{p{\textwidth}}{
    The coordinates and visual magnitudes are taken from the Simbad database ({\it http://simbad.u-strasbg.fr/simbad/}). The periods are adopted from Norton et al. 2011.}
}

\MakeTable{llcc}{6cm}{The eclipse times measured from our data.}
{
  \hline
    Nickname & $\rm T_I$ [HJD] & $\rm T_{II}$ [HJD] \\
    \hline
    J022726  & 2457009.6198    & 2457009.7257       \\
    J040615  & 2457013.6276    & 2457013.7392       \\
    J121906  & 2456768.6441    & 2456768.5299       \\
    J232610  & 2457274.9329    & 2457274.8167       \\
  \hline
}

While Koen et al. (2016) do not provide measurement errors for individual observations, they estimate that these errors for the four stars studied in this work are under 0.02 mag. We adopt this constant as the standard error of the observations when calculating the $\chi^2$ metrics.

\section{The q-Search}

We will not repeat the description of the initial q-search, which is given in detail in P1. In what follows, we refer to this step of the analysis as ``the randomized q-search''. Its results are the starting point for the current work. They consist of 100 models initialized with random parameters and then optimized to best fit the observations for each mass ratio in the range from 0.01 to 1.00 with the step of 0.01; that's 10,000 models for each of the two possible configurations: the A-type (the more massive star is also the hotter one) and the W-type (the more massive star is colder than the companion). When the $\chi_{\nu}^2$ of the best-fitting model in each mass ratio bin is plotted against mass ratio, we get a ``q-search curve'' (see Fig. 1). For all 29 stars in the dataset of Koen et al. 2016, it is possible to prejudge the A/W configuration of the system just by looking at these q-search curves. One configuration is always superior to the other (in the sense of achieving better fits to the observations). In this work, we select the better of the two configurations and disregard the other. According to this indirect criterion, J022726 is a W-type star and all the others in the current sample are A-types.

We made the selection of stars to study further despite the lack of a visible total eclipse on the basis of these q-search curves. In Fig. 1, we compare the q-search curves of J00437 (one of the stars from P1, which displays a clear total eclipse), J022726 (the star in the current sample with the ``best'' q-search curve) and 1SWASP J212808.86+151622.0 (an example of a ``bad'' q-search curve for a star from the same dataset, which we didn't include in neither P1 nor this study). A line is drawn on each of these plots at arbitrary coordinates to aid the eye in recognizing that the first two curves have discernible minima, while the third is monotonous, with the best fit achieved at mass ratio of 1. In contact binaries of the W UMa type, mass ratios are typically around 0.3 and rarely exceed 0.5, so a mass ratio as high as this would be highly unusual and therefore suspect. We show in Section 4 that the estimated mass ratios for the four stars investigated in the present work lie within the range expected for W UMa stars.

\begin{figure}[htb]
  \includegraphics[width=0.5\textwidth]{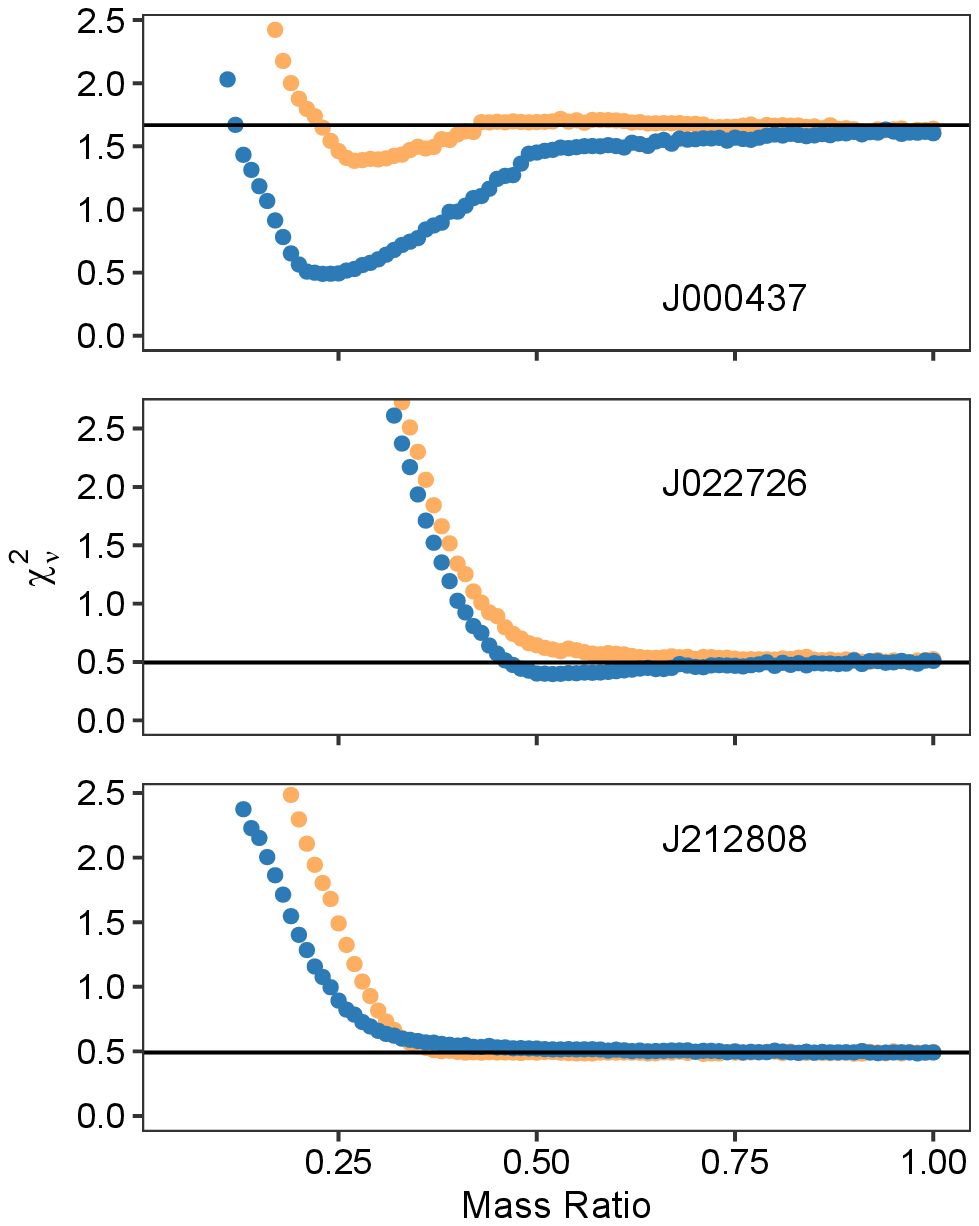}
  \includegraphics[width=0.5\textwidth]{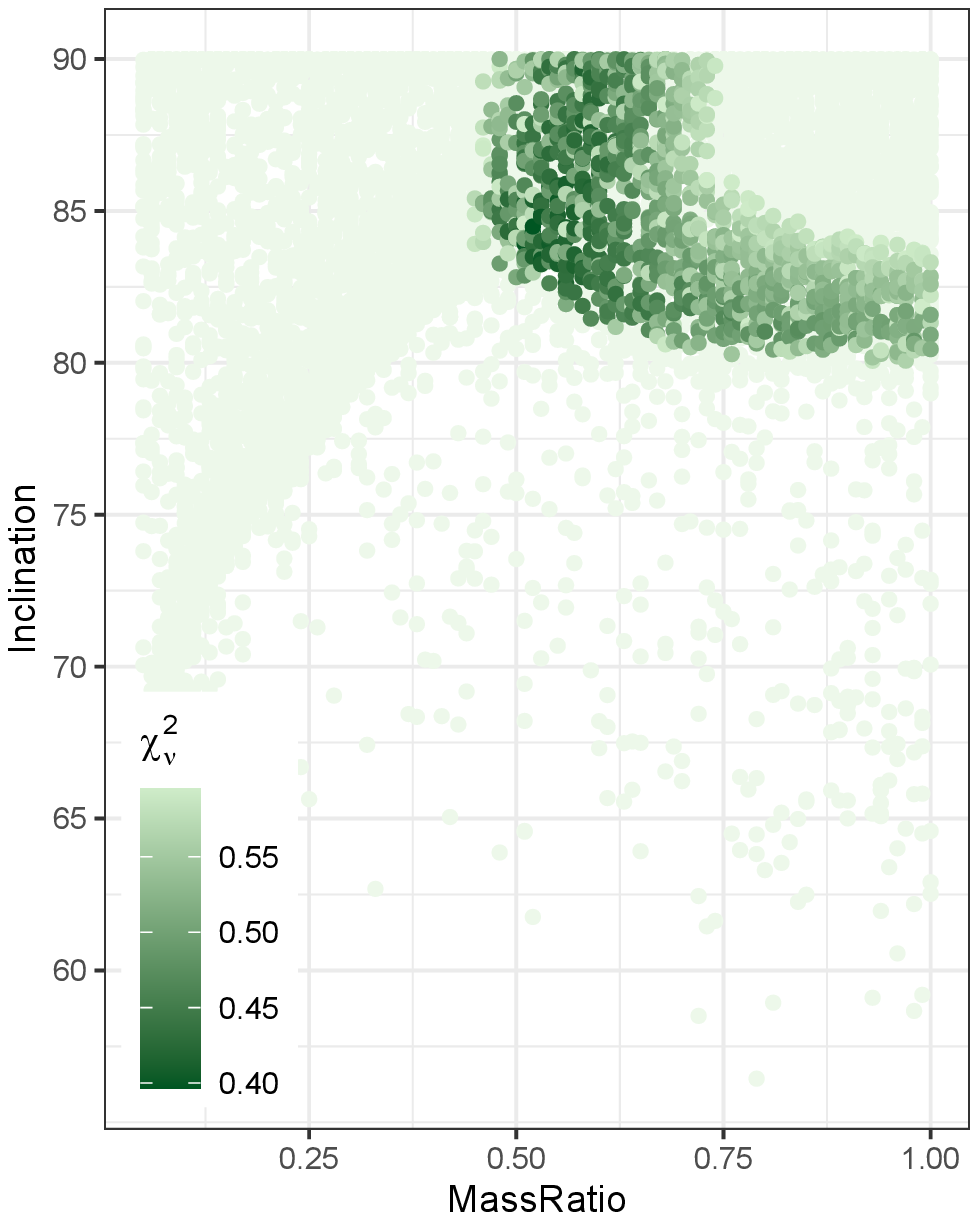}
  \FigCap{Left: A comparison of q-search curves for a contact binary with a clearly visible total eclipse (1SWASP J000437.82+033301.2, studied in P1), one of the binaries from the current study (J022726), and another binary from the same series of observations (1SWASP J212808.86+151622.0), for which the q-search fails to constrain the mass ratio. Right: The results of the one-dimensional, randomized q-search for J022726 in the $q-i$ plane. Darker points correspond to better-fitting models (lower values of $\chi_{\nu}^2$). Compare with middle panel on the left.}
\end{figure}

In randomized q-search performed in P1, all relevant model parameters apart from the mass ratio are adjusted to get an optimal fit, and tabulated for all trial models. The most important parameter in terms of constraining all subsequent results is the orbital inclination. Plotting the inclination against the mass ratio can identify the region of the parameter space that contains the globally optimal model. Fig. 1 shows such a plot for J022726, based on the randomized q-search. Each point is a trial model; the colors are set up so that all the models with $\chi_{\nu}^2 > 1.25~min(\chi_{\nu}^2)$ are a very pale green, enhancing the visibility of the ``good'' candidates. They are clearly confined to the region roughly between mass ratio 0.45 and 0.65 and inclinations between 80 and 90 degrees.

Based on such plots, we select a region in the $q-i$ plane to examine in greater detail using the ``grid q-search''. We generate new initial models along a $q-i$ grid with the step in the $q$ direction of 0.001 (an order of magnitude finer than in the randomized q-search) and in the $i$ direction of 0.1 degree. In the case of J022726, the selected region is $0.48 \le q \le 0.68$ and $80^{\circ} \le i \le 90^{\circ}$ and contains around 20 thousand candidate models. For J040615 and J232610, where a larger region had to be considered, the size of the grid is closer to 60 and 70 thousands, respectively. The mass ratio and the inclination in these models were kept fixed to the grid values; the other parameters were initialized by copying the best-fitting model in the corresponding mass ratio bin from the randomized q-search, and then optimizing to best fit the observations.

The results of the grid q-search are shown in Fig. 2.

\begin{figure}[htb]
  \includegraphics[width=\textwidth]{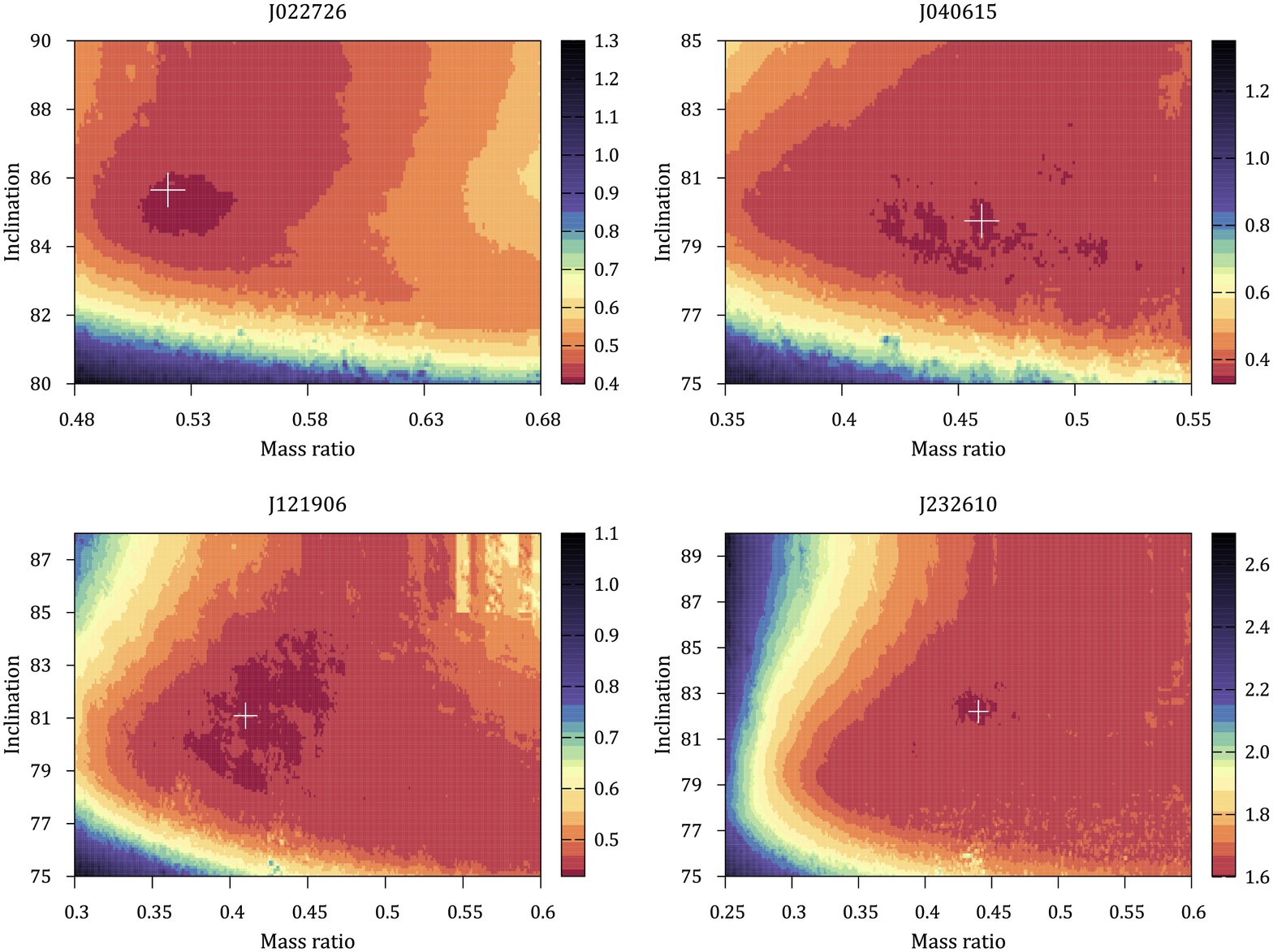}
  \FigCap{The results of the 2D q-search for our stars. Each pixel represents a trial model. The goodness of fit ($\chi_{\nu}^2$) is color-coded; the best models with lowest $\chi_{\nu}^2$ are in the central part of each plot, and indicated by a dark red color. The final solution is marked with the white cross.}
\end{figure}

\section{Detailed Modeling}

The best-fitting model resulting from the 2D q-search described above is taken as the starting point for detailed modeling. Here and throughout this paper, light curve modeling is done with the program introduced by Djura\v sevi\' c (1992) and Djura\v sevi\' c et al. (1998), as described in P1.

In this step of the analysis, we examine the details of the light curve and add spots to the model to match any out-of-eclipse asymmetries. The mass ratio ($q$), orbital inclination ($i$), passband-dependent third light contributions ($\ell_3$), size (parametrized with the filling factor, $F$, which is equal to the ratio of the critical Roche potential and the Roche potential of the stellar surface), the temperature of the primary (secondary) component in the W (A) configuration, the spot parameters and the phase and magnitude shifts are treated as free parameters. For the albedos ($A_1$ and $A_2$) and gravity darkening exponents ($\beta_1$ and $\beta_2$), we adopt the theoretical values appropriate for each component according to its temperature (von Zeipel 1924; Lucy 1967; Ruci{\'n}ski 1969). For each system, we assign the temperature estimated by Koen et al. (2016) to the component eclipsed in the deeper minimum, while the temperature of the other star is adjusted as a free parameter. The spots are kept fixed at the equator.

The final model parameters of our four stars are given in Table 3. To estimate the errors, we perform random sampling of the parameter space around the $\chi^2$ minimum (achieved with the final model), in a range defined by the increase of the $\chi^2$ by 1. This provides the 1$\sigma$ error estimate for each fitted parameter. Note that the temperature of one component is always fixed to a referent value. Due to this, the reported errors are necessarily optimistic.

The observations are plotted together with the synthetic light curves corresponding to these models in Fig. 3, and 3D representations of the binaries are shown in Fig. 4. Using these results, we estimate the absolute stellar parameters of our targets in the next section.

\MakeTable{lrrrr}{12.5cm}{The model parameters for the studied stars.}
{
  \hline
  Quantity              & J022726   & J040615   & J121906   & J232610   \\
  \hline
  $q$                   & 0.52(4)   & 0.46(6)   & 0.41(3)   & 0.44(6)   \\
  $i [^{\circ}]$        & 86(2)     & 80(2)     & 81(3)     & 82(2)     \\
  $T_1 [K]$             & 3980(20)  & 4840      & 4840      & 4840      \\
  $T_2 [K]$             & 4050      & 4680(20)  & 4740(30)  & 4900(20)  \\
  $F_{1,2}$             & 1.020(7)  & 1.013(2)  & 1.027(6)  & 1.025(3)  \\
  $\ell_3 (B)$          & 0.00(2)   & 0.07(3)   & 0.07(6)    & 0.09(3)  \\
  $\ell_3 (V)$          & 0.03(3)   & 0.06(3)   & 0.10(7)    & 0.11(3)  \\
  $\ell_3 (R)$          & 0.07(3)   & 0.07(3)   & 0.12(6)    & 0.13(4)  \\
  $\ell_3 (I)$          & 0.15(2)   & 0.09(3)   & 0.15(6)    & 0.16(3)  \\  
  \hline
  $r_1 [a_{orb}]$       & 0.4507    & 0.4569    & 0.4761    & 0.4695    \\
  $r_2 [a_{orb}]$       & 0.3354    & 0.3230    & 0.3222    & 0.3258    \\  
  $\Omega_{1,2}$        & 2.8567    & 2.7716    & 2.6357    & 2.6904    \\
  $\Omega_{in}$         & 2.9053    & 2.8040    & 2.6975    & 2.7496    \\
  $\Omega_{out}$        & 2.5985    & 2.5254    & 2.4482    & 2.4860    \\
  $f_{over} [\%]$       & 15.83     & 11.64     & 24.79     & 22.46     \\
  \hline                                                                  
  Spot                  & -         & Primary   & Primary   & Primary   \\
  \hline                                                                  
  $T_{spot}/T_{star}$   & -         & 0.9(1)    & 0.9(2)    & 0.9(2)    \\
  $\sigma$              & -         & 16(2)     & 14(1)     & 21(2)     \\
  $\lambda$             & -         & 192(4)    & 107(4)    & 95(3)     \\
  $\varphi$             & -         & 0.00      & 0.0       & 0.00      \\
  \hline                                                                  
  Point count       & 168    & 223    & 246       & 357       \\
  Parameter count             & 10     & 13     & 13        & 13        \\
  Degrees of freedom ($\nu$)  & 158    & 210    & 233       & 344       \\
  $\chi_{\nu}^2$    & 0.4244 & 0.2601 & 0.1948    & 0.4806    \\
  \hline
  %\multicolumn{5}{p{6cm}}{Notes.}
}

\begin{figure}[htb]
  \includegraphics[height=0.9\textheight]{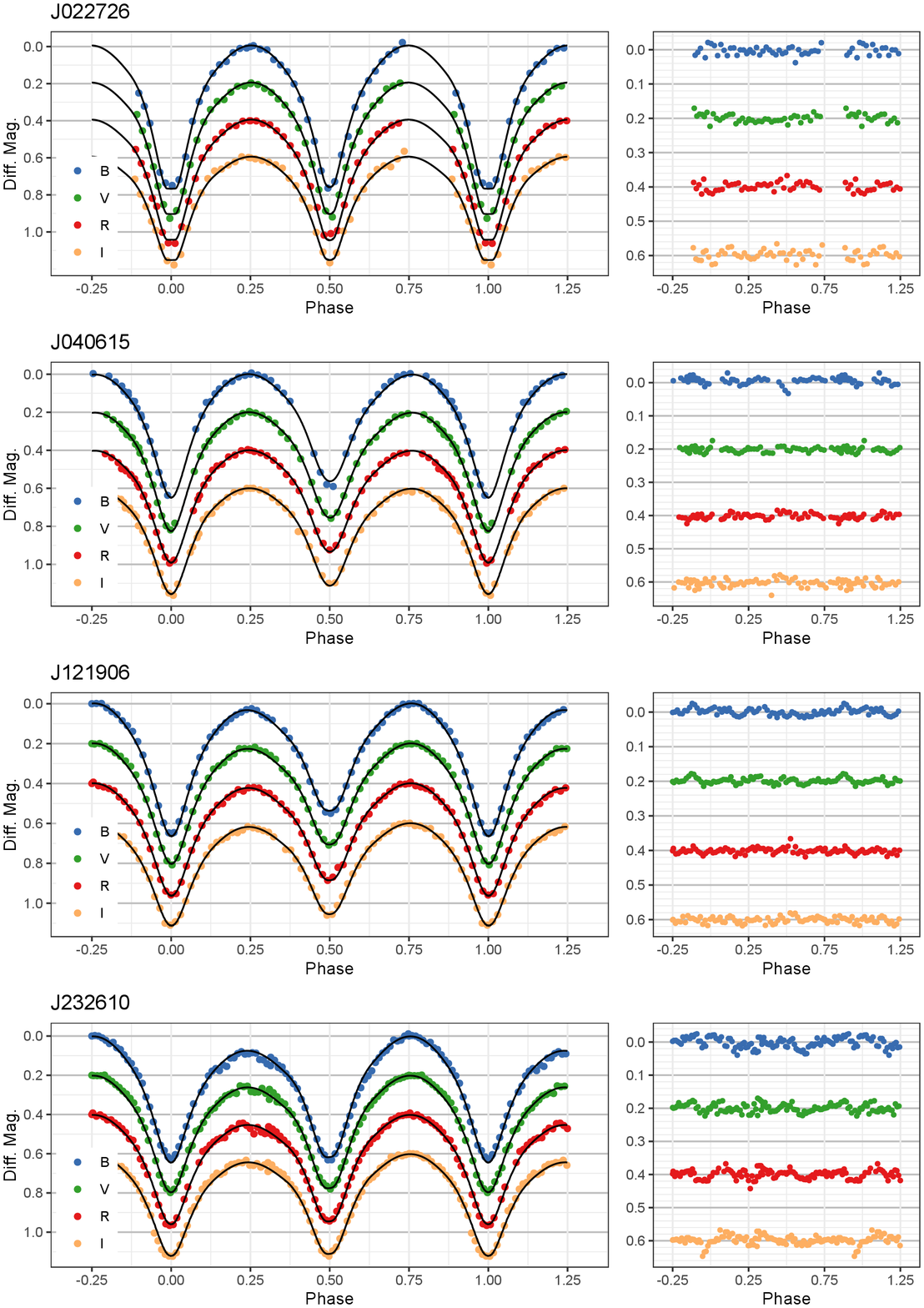}
  \FigCap{The observed and synthetic light curves of our stars (left), with residuals (right) arbitrarily shifted along the magnitude axis for clarity.}
\end{figure}

\begin{figure}[htb]
  \includegraphics[width=\textwidth]{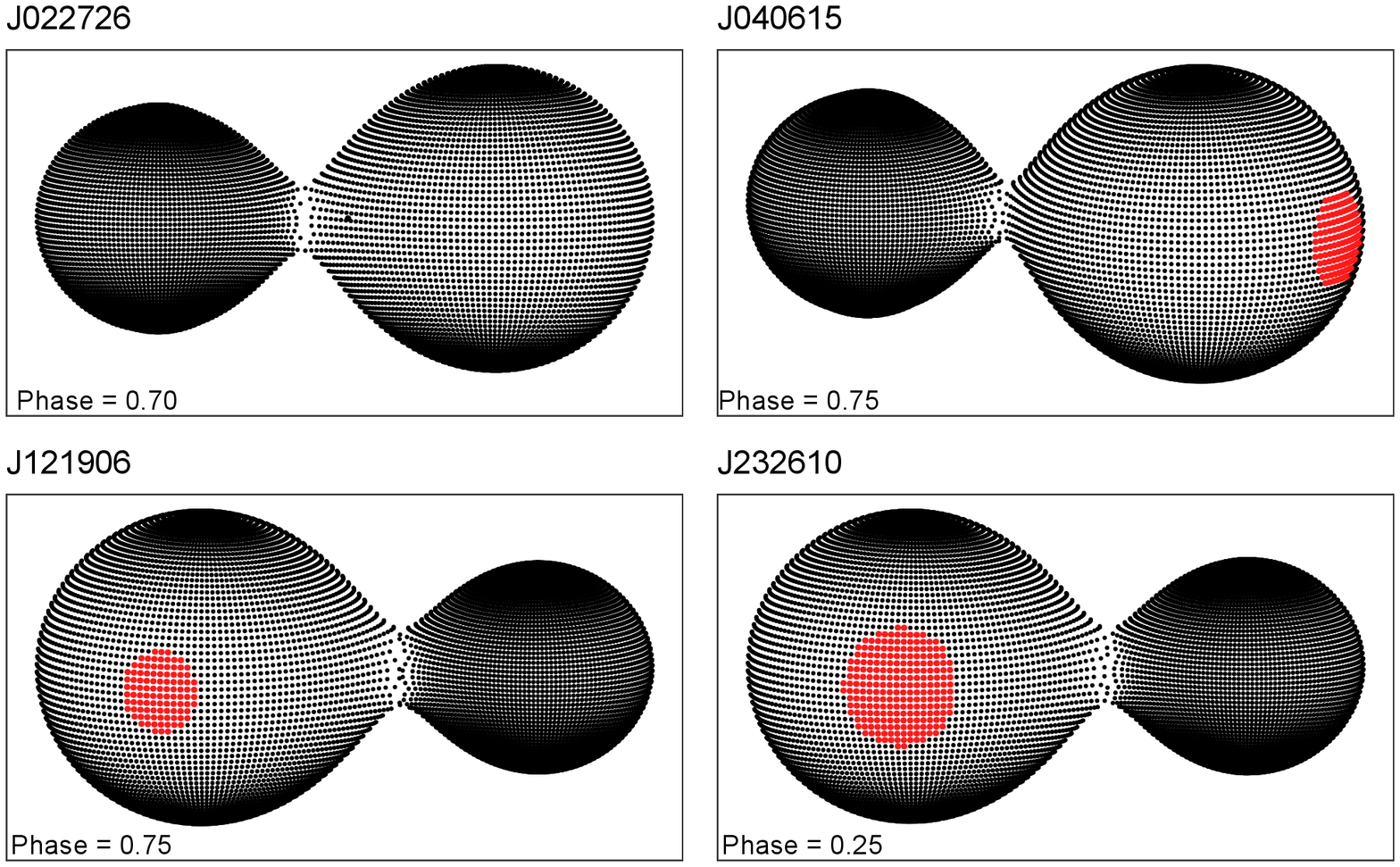}
  \FigCap{The 3D representations of the models for our stars. The spots are marked in a different color.}
\end{figure}

Note that, while the initial model for J232610 indicated the A-type configuration, the final model rather implies the W-type configuration, with a secondary of a slightly higher temperature than the primary. This is likely a consequence of the addition of a cool spot on the primary that has the effect of decreasing its brightness, which is then compensated during model optimization by increasing the temperature of the secondary. Clearly, this is a marginal case, where radial velocity time-series would be needed for a definite classification. For now, we classify J232610 as a W-type contact binary according to the final model.

\section{Absolute Parameters and Evolutionary Status}

To estimate the masses, radii and luminosities of the components of our four stars in solar units, we adopt an approach similar as in P1. Under the assumption that the more massive primary has not evolved far from the main sequence, we interpolate its mass from the tabulations derived by Eker at al. (2018) based on a large sample of well-studied, detached binaries, according to the temperature estimated by Koen et al. (2016). The secondary mass can then be calculated from the mass ratio, and the separation from the third Kepler law. For the referent temperature, we adopt the uncertainty of 100 K reported by Koen et al. (2016). The uncertainty in the primary mass is then estimated as the range of masses from the Eker et al. (2018) tabulations corresponding to the error of the primary temperature. The results are given in Table 4.

\MakeTable{lrrrr}{12.5cm}{The absolute parameters of our stars.}
{
  \hline
  Quantity          & J022726   & J040615   & J121906   & J232610   \\ 
  \hline                                                       
  $a [R_{\odot}]  $ & 1.45(2)   & 1.65(4)   & 1.65(3)   & 1.68(4)   \\ 
  $M_1 [M_{\odot}]$ & 0.609(7)  & 0.84(3)   & 0.84(3)   & 0.84(3)   \\ 
  $M_2 [M_{\odot}]$ & 0.32(3)   & 0.38(6)   & 0.34(4)   & 0.37(6)   \\ 
  $R_1 [R_{\odot}]$ & 0.65(2)   & 0.75(2)   & 0.79(2)   & 0.79(3)   \\ 
  $R_2 [R_{\odot}]$ & 0.48(1)   & 0.53(2)   & 0.53(2)   & 0.55(2)   \\ 
  $T_1 [K]$         & 3980(20)  & 4840(100) & 4840(100) & 4840(100) \\ 
  $T_2 [K]$         & 4050(100) & 4680(20)  & 4740(30)  & 4900(20)  \\ 
  $L_1 [L_{\odot}]$ & 0.095(6)  & 0.28(4)   & 0.30(4)   & 0.30(5)   \\ 
  $L_2 [L_{\odot}]$ & 0.057(8)  & 0.122(8)  & 0.128(9)  & 0.16(2)   \\ 
  $\log(g)_1$       & 4.59(3)   & 4.60(4)   & 4.57(3)   & 4.57(4)   \\ 
  $\log(g)_2$       & 4.57(6)   & 4.57(9)   & 4.52(7)   & 4.5(1)    \\   
  \hline                                                       
}

Judging by their absolute parameters, our four targets are typical short-period W UMa stars, with sub-solar, unevolved primaries and secondaries that are undersized and udnerluminous compared to main-sequence stars of similar mass and temperature, in consequence of matter and energy transfer through the common envelope. A comparison with a sample of similar systems with periods shorter than 0.25 days (listed in Table 5), demonstrates this further. Fig. 5 shows this sample together with our targets on the HR diagram. The main sequence, extracted from the MIST model archive (Dotter 2016; Choi et al. 2016) is indicated as well. The primaries of J040615, J121906 and J232610, having the same temperature, form a tight group, while the secondaries are scattered over a greater range. We take the fact that our stars occupy the same region of the HR diagram as the sample of totally eclipsing short-period W UMa binaries from Table 5 as additional evidence that the absolute parameters estimated from our two-dimensional q-search are fairly reliable.

\MakeTable{lrrrrrrrrrrr}{12.5cm}{A sample of totally-eclipsing W UMa binaries with periods shorter than 0.25 days.}
{
\hline
Star                         &  $P$     &  $q$   &  $M_1$ &  $M_2$ &  $R_1$ &  $R_2$ &  $T_1$ &  $T_2$ &  $L_1$  &  $L_2$  &  Ref.  \\
\hline                                                       
SDSS J012119.10-001949.9     &  0.2052  &  0.50  &  0.51  &  0.26  &  0.61  &  0.45  &  3840  &  3812  &  0.07  &  0.04  & $[1]$  \\
2MASS J21042404+0731381      &  0.2091  &  0.32  &  0.59  &  0.19  &  0.67  &  0.40  &  4220  &  4450  &  0.13  &  0.06  & $[2]$  \\
NSVS 7179685                 &  0.2097  &  0.47  &  0.65  &  0.30  &  0.67  &  0.48  &  3979  &  4100  &  0.10  &  0.06  & $[3]$  \\
1SWASP J080150.03+471433.8   &  0.2175  &  0.43  &  0.72  &  0.32  &  0.71  &  0.49  &  4685  &  4696  &  0.23  &  0.11  & $[3]$  \\
CC Com                       &  0.2207  &  0.53  &  0.72  &  0.38  &  0.71  &  0.53  &  4200  &  4300  &  0.14  &  0.08  & $[4]$  \\
1SWASP J074658.62+224448.5   &  0.2208  &  0.35  &  0.79  &  0.28  &  0.80  &  0.52  &  4543  &  4717  &  0.24  &  0.12  & $[5]$  \\
NSVS 2175434                 &  0.2209  &  0.33  &  0.81  &  0.27  &  0.80  &  0.51  &  4898  &  4903  &  0.33  &  0.13  & $[5]$  \\
1SWASP J052926.88+461147.5   &  0.2266  &  0.41  &  0.80  &  0.33  &  0.77  &  0.52  &  5077  &  5071  &  0.36  &  0.16  & $[6]$  \\
1SWASP J093010.78+533859.5   &  0.2277  &  0.40  &  0.86  &  0.34  &  0.79  &  0.52  &  4700  &  4700  &  0.27  &  0.12  & $[7]$  \\
1SWASP J212454.61+203030.8   &  0.2278  &  0.44  &  0.76  &  0.33  &  0.75  &  0.52  &  4840  &  4810  &  0.28  &  0.13  & $[2]$  \\
1SWASP J044132.96+440613.7   &  0.2281  &  0.64  &  0.70  &  0.45  &  0.72  &  0.60  &  4003  &  3858  &  0.12  &  0.07  & $[6]$  \\
2MASS J21031997+0209339      &  0.2286  &  0.48  &  0.51  &  0.24  &  0.65  &  0.47  &  3927  &  4050  &  0.09  &  0.05  & $[2]$  \\
1SWASP J050904.45-074144.4   &  0.2296  &  0.44  &  0.76  &  0.33  &  0.75  &  0.52  &  4840  &  4933  &  0.28  &  0.14  & $[2]$  \\
V1009 Per                    &  0.2341  &  0.36  &  0.87  &  0.31  &  0.86  &  0.47  &  5280  &  5253  &  0.52  &  0.15  & $[8]$  \\
YZ Phe                       &  0.2347  &  0.38  &  0.74  &  0.28  &  0.76  &  0.49  &  4658  &  4908  &  0.24  &  0.12  & $[9]$  \\
1SWASP J195900.31-252723.1   &  0.2381  &  0.51  &  0.81  &  0.41  &  0.78  &  0.57  &  5027  &  5170  &  0.35  &  0.21  & $[2]$  \\
1SWASP J064501.21+342154.9   &  0.2486  &  0.48  &  0.70  &  0.30  &  0.76  &  0.55  &  4590  &  4720  &  0.23  &  0.13  & $[10]$ \\
\hline                                                       
\multicolumn{11}{p{12.5cm}}{
  $[1]$ Jiang et al. (2015); 
  $[2]$ Latkovi{\' c} \& {\v C}eki (2021);
  $[3]$ Dimitrov \& Kjurkchieva (2015);
  $[4]$ K{\"o}se et al. (2011);
  $[5]$ Kjurkchieva et al. (2018a);
  $[6]$ Kjurkchieva et al. (2018b);
  $[7]$ Lohr et al. (2015);
  $[8]$ Michel et al. (2019);
  $[9]$ Sarotsakulchai et al. (2019);
  $[10]$ Djura{\v{s}}evi{\'c} et al. (2016).
  }
}

\begin{figure}[htb]
  \includegraphics[width=\textwidth]{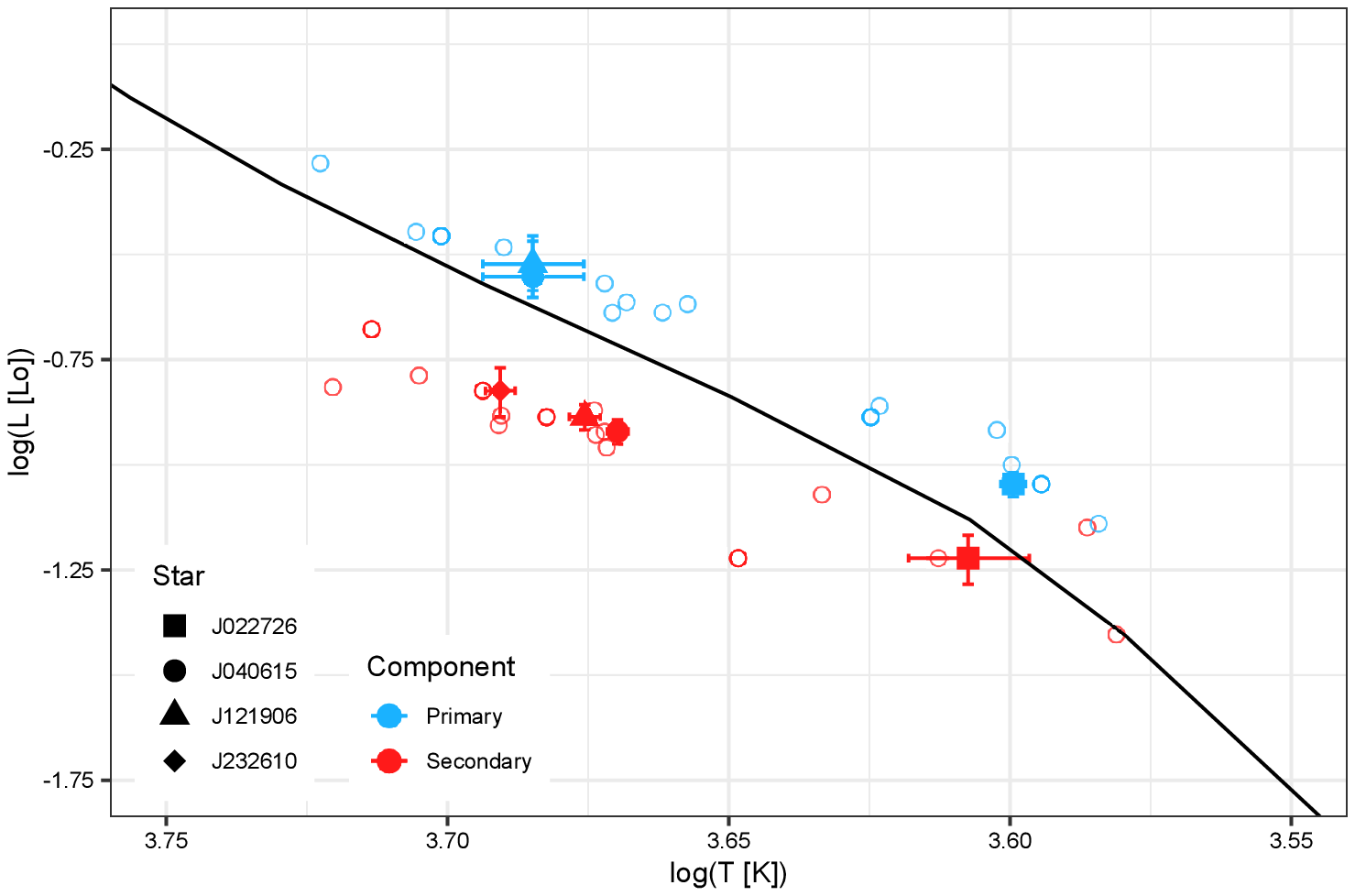}
  \FigCap{Our four stars on the HR diagram. The primaries are colored blue, and the secondaries, red. Objects from Table 5 are also shown as smaller, empty circles. The solid line indicates the main sequence extracted from the MIST model archive (Dotter 2016; Choi et al. 2016).}
\end{figure}

Among our targets, J022726 is the only star that was studied prior to this work. Liu et al. (2015) analyzed its multicolor CCD light curves (displaying a convincing totality in the primary minimum) and derived its global parameters with similar methods as those we used here. Overall, our results are fairly consistent. We obtain a higher mass ratio (0.52 in ours vs. 0.46 in their study), larger degree of contact (16 vs. 10\%) and slightly higher masses (0.61 vs. 0.54 $M_{\odot}$ for the primary and 0.32 vs. 0.25 $M_{\odot}$ for the secondary component). Liu et al. (2015) found it necessary to include a small dark spot visible around orbital phase 0.75 in their model, while in our study, this addition isn't required to reproduce the observations. Long-term variability due to the appearance, disappearance or migration of spots is common-place among W UMa binaries, whose late-type components are expected to exhibit magnetic activity, so this discrepancy is not surprising.

\section{Concluding Remarks}

W UMa stars are interesting objects whose properties, evolution and membership in multiple stellar systems are not fully understood even after decades of research. As the most common class of eclipsing binaries, they make a significant fraction of variable stars observed by space telescopes and ground-based surveys. In the context of the current and future deluge of light curves, it is important to develop techniques that can be used to extract the physical properties of stars from this data even without the spectroscopic follow-up. The q-search is one such technique, but the limits of its reliability are not well-known. 

In this work, we performed a detailed, 2-dimensional q-search for four W UMa binaries near the period cutoff, using extensive grids of binary system models (with up to 70000 models per grid). We showed that the mass ratios and inclinations of our targets are confined to closed regions of the parameter space that can be assumed to contain the best possible model (the global minimum of the $\chi_{\nu}^2$). The resulting absolute parameters can therefore be considered as robust estimates and make a significant contribution to the still small sample of ultra-short-period late-type contact binaries.

% Acknowledgements

\Acknow{
  This research was funded by the Ministry of Education, Science and Technological Development of Republic of Serbia (contract No. 451-03-68/2022-14/200002). The authors gratefully acknowledge the use of the Simbad database ({\it http://simbad.u-strasbg.fr/simbad/}), operated at the CDS, Strasbourg, France, and NASA's Astrophysics Data System Bibliographic Services ({\it http://adsabs.harvard.edu/})
}

% References

\end{document}